\RequirePackage{fix-cm}
\documentclass[smallextended,natbib]{svjour3}       
\smartqed  
\usepackage[misc,geometry]{ifsym} 
\usepackage{graphicx,amsmath,amssymb}
\usepackage[dvipsnames]{xcolor}
\newcommand{\m}[1]{\langle #1 \rangle}
\newcommand{\E}{{\rm I\!E}}
\newcommand{\Cov}{\mathrm{Cov}}
\begin{document}


\title{Measuring Tree Balance with Normalized Tree Area}
%



\author{T. Ara\'ujo Lima \and Flavia M. D. Marquitti \and Marcus A. M. de Aguiar 
}


\institute{
 \begin{acknowledgements}
  This work was partly funded by the Brazilian agencies: TAL was supported by grant \#158002/2018-0 (CNPq). FMDM was supported by grant \#2015/11985-3 (FAPESP) and Capes. MAMA was supported by grant \#301082/2019-7 (CNPq), \#2016/01343-7 (ICTP-SAIFR FAPESP) and \#2019/20271 (FAPESP). 
 \end{acknowledgements}
              T. Ara\'ujo Lima (\Letter) \at
              \email{tiagoapl@ifi.unicamp.br}           
           \and
           Flavia M. D. Marquitti \at
           \email{flamarquitti@gmail.com}
           \and
           Marcus A. M. de Aguiar \at
              \email{aguiar@ifi.unicamp.br}\\
              \newline
              Instituto de F\'{i}sica ‘Gleb Wataghin’, Universidade Estadual de Campinas, Unicamp, 13083-970 Campinas, SP, Brazil
}

\date{Received: date / Accepted: date}

\maketitle

\begin{abstract}

The study of species organization and their clustering by genetic or phenotypic similarity is carried out with the tools of  phylogenetic trees. An important structural property of phylogenetic trees is the balance, which measures how taxa are distributed among clades. Tree balance can be measured using indices such as the Sackin ($S$) and the Total Cophenetic ($\Phi$),  which are based on the distance between nodes of the tree and its root. Here, we propose a new metric for tree balance, $\bar{d}$, the Area per Pair (APP) of the tree, which is a re-scaled version of the so called tree area. We compute $\bar{d}$ for the rooted caterpillar and maximally balanced trees and  we also obtain exact formulas for its expected value and variance under the Yule model. The variance of APP for Yule trees has the remarkable property of converging to an asymptotic constant value for large trees. We compare the Sackin, Total Cophenetic and APP indices for hundreds of empirical phylogenies and show that APP represents the observed distribution of tree balances better than the two other metrics.

\end{abstract}

\section{Introduction}
\label{intro}

\ \ \ \ Several analytical and computational tools have been recently developed  to understand and characterize phylogenetic trees \citep{fel:2004}. These studies are motivated by the belief that structural properties of trees reflect their evolutionary history and were shaped by processes such as geographic modes of diversification and by rates of speciation and extinction events.  \citep{nee:1992,moo:1997,mor:2014,cab:2017,cos:2019}. Extracting evolutionary information from a tree, however, is not a simple task. Here we consider rooted binary tree graphs where the leaves (tips) denote living (extant) species and the internal nodes represent the point in time when a speciation event occurred. It is appropriate to call the internal nodes {\it ancient species} that branched into the living ones. If $n$ is the number of leaves, then the total number of species (linving and ancient) in the tree is $N=2n-1$.\\

An important feature of binary trees is their balance. In simple terms, a tree is balanced if it has equal numbers of leaves emanating from both branches of its bifurcations (internal nodes) . By contrast, imbalance is the opposite property, unequal numbers of leaves arising from each branch \citep{sha:1990}. As the numbers of species in studies of phylogenetic trees has grown, so has the interest in using the shapes of these trees to test hypothesis about evolution. For example, if a few lineages give rise to most of the descendant species because they have acquired an important adaptation, this should be visible in the imbalance of the resulting tree \citep{fel:2004}. Simulations and comparisons with real phylogenies were performed by \cite{cos:2019}, showing that in a neutral model of evolution balanced trees evolve more frequently if speciation is sympatric (when individuals inhabit the same geographic region). Populations displaying geographical structures (parapatry and allopatry), on the other hand, give rise to more unbalanced trees.\\

Measures such as the index of Sackin \citep{sac:1972,sha:1990,fis:2019}, Colless \citep{col:1982,cor:2020}, and Total Cophenetic indices \citep{mir:2013,car:2013} quantify tree balance. Here we focus on the Sackin and Total Cophenetic, as the relate directly to the new index we propose. The index of Sackin is calculated as the sum of all  distances between the leaves ($n$ terms) to the root node of the binary tree. The Total Cophenetic index takes all the distances from the most recent common ancestor (MRCA) of each pair of leaves to the root node and sums them up (${n \choose 2}$ terms). MRCA is the node that represents the closest internal node from whom the pair of leaves descend directly.\\

In this paper, we propose a new topological balance metric based on the distances between tips, which we call $\bar{d}_n$, the Area Per Pair (APP). Distances between tips are natural quantities, as they can be related to other distance measures, such as genetic and phenotypic differences between species or individuals. In sec. \ref{area} we  define the APP index and relate it with the Sackin and Total Cophenetic indices. Next  we calculate the APP for the special cases of maximally balanced and fully unbalanced trees. In sec. \ref{yule} we obtain exact formulas for its expected value and variance under the Yule model and in sec. \ref{real} we compare Sackin, Total Cophenetic and APP indices for a large set of empirical trees. Sec. \ref{conc} summarizes our conclusions.

\begin{figure*}[!htpb]
 \centering
 \includegraphics[width=0.7\textwidth]{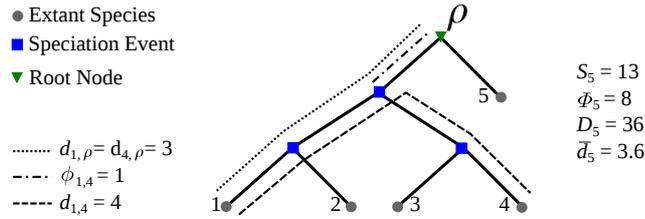}
 \caption{Distances $d_{i,j}$ and $\phi_{i,j}$ in a hipothetical binary tree. Check metrics information in section \ref{intro}. The green triangle represents the root $\rho$, nodes labeled from 1 to 4 are tips (extant species in phylogenetic trees) and blue squares are internal nodes, representing bifurcations (speciation events in phylogenetic trees).}
 \label{scheme}
\end{figure*}

\section{The Area Per Pair (APP) Index}
\label{area}

\ \ \ \ In phylogenetic trees, leaves (or tips) correspond to the living (or extant) species.  In this paper we shall consider only rooted binary trees, where each node is either a leaf, an internal node  (giving rise to two new branches), or the root. The distance between leaves $i$ and $j$ is defined as the number of edges that connect the corresponding tips in the tree, also called genealogic distance \citep{ste:2001,mul:2011}:
\begin{equation}
 \label{wij}
 d_{i,j} = d_{i,\rho} + d_{j,\rho} -2\phi_{i,j},
\end{equation}
where $\rho$ is the root node, $d_{i,\rho}$ is the distance between $i$ and the root and $\phi_{i,j}$ is the cophenetic value of $(i,j)$, the distance between their most recent common ancestor (MRCA) and the root. Figure \ref{scheme} shows a simple example illustrating these quantities. The Area Per Pair Index (APP), $\bar{d}_n$, is defined as the average of distances between all pairs of tips in the binary tree :
\begin{equation}
 \label{wbar}
 \bar{d}_n = \frac{2}{n(n-1)}\sum_{\m{i,j}} d_{i,j}
\end{equation}
where $n$ is the number of leaves in the tree. APP is  a re-scaled version of the total tree area \citep{mul:2011,mir:2013} 
\begin{equation}
 \label{Dnwbar}
 D_n = \sum_{\m{i,j}} d_{i,j} = \frac{n(n-1)}{2}\bar{d}_n
\end{equation}
The total area, in turn, is related to the Sackin index --$S_n$, \citep{sac:1972,sha:1990,fis:2019} and Total Cophenetic value, $\Phi_n$, defined by \cite{mir:2013}:
\begin{equation}
 \label{Sn}
 S_n = \sum_{i=1}^n d_{i,\rho},
\end{equation}
and
\begin{equation}
 \label{Phin}
 \Phi_n = \sum_{\m{i,j}} \phi_{i,j}.
\end{equation}
Summing over all pairs of leaves in eq. (\ref{wij}) we obtain:
\begin{equation*}
 \sum_{i=1}^{n-1}\sum_{j=i+1}^n d_{i,j} = \sum_{i=1}^{n-1}\sum_{j=i+1}^n (d_{i,\rho} + d_{j,\rho}) -2\sum_{i=1}^{n-1}\sum_{j=i+1}^n \phi_{i,j}.
\end{equation*}
or
\begin{equation}
 \label{DSPhi}
 D_n = (n-1)S_n - 2\Phi_n.
\end{equation}
Comparing eqs. (\ref{Dnwbar}) and (\ref{DSPhi}) we can write $\bar{d}_n$ as function of $S_n$ and $\Phi_n$:
\begin{equation}
 \label{wSPhi}
 \bar{d}_n = \frac{2}{n}S_n - \frac{4}{n(n-1)}\Phi_n.
\end{equation}
In the next section we will calculate the APP index for two special cases of binary trees, namely, the fully unbalanced (rooted caterpillar) and maximally balanced trees. For the Sackin and Total Cophenetic indices these cases correspond to the maximum and minimum values attained for rooted binary trees \citep{fis:2019,mir:2013}.

\section{APP Index for some Special Trees}
\label{spetree}

\subsection{Fully Unbalanced Trees}
\label{asym}

\begin{figure*}[!htpb]
 \centering
 \includegraphics[width=0.75\textwidth]{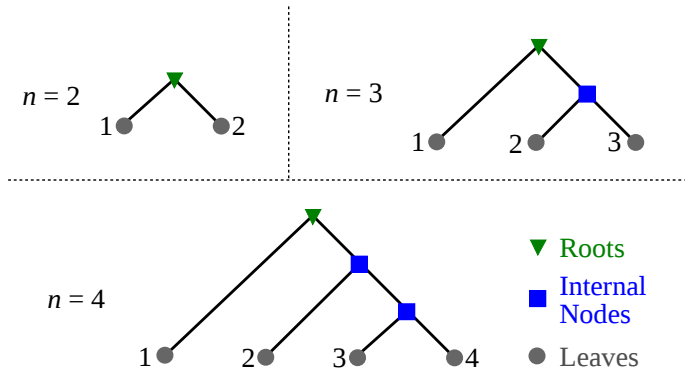}
 \caption{Fully unbalanced trees with 2 to 4 leaves. The green triangle represents the root node $\rho$, nodes labaled with numbers are extant species (tips) and blue squares are internal nodes.}
 \label{trees_unbal}
\end{figure*}

\ \ \ \ Fully unbalanced trees are called caterpillars, characterized by all bifurcations occurring in only one branch, from the same leaf and internal node. Figure \ref{trees_unbal} shows examples of these trees. Exact expressions for the Sackin ($S_n$) and Total Cophenetic ($\Phi_n$) indices are known for these cases \citep{fis:2019,mir:2013}:
\begin{equation}
 \label{Sunb}
 S_n^{(unb)} = \frac{(n-1)(n+2)}{2},
\end{equation}
\begin{equation}
 \label{Phiunb}
 \Phi_n^{(unb)} = \frac{n}{6}(n-1)(n-2).
\end{equation}
Using these results we can write eq. (\ref{wSPhi}) as
\begin{eqnarray}
\label{wunb}
 \bar{d}_n^{(unb)} &=& \frac{2}{n} \, \frac{(n-1)(n+2)}{2} - \frac{4}{n(n-1)}\cdot\frac{n}{6}(n-1)(n-2) \nonumber \\
  &=& \frac{n+7}{3} - \frac{2}{n}, \quad \text{for }n>1,
\end{eqnarray}
where $\bar{d}_1^{(unb)}=0$. For fixed tree size -- $n$, the Sackin  and Total Cophenetic indices have maximal values for binary caterpillar trees. For the APP index the value of $ \bar{d}_n^{(unb)}$ is not always maximal, as we will discuss later at the end of section \ref{bal}. We note that, for large $n$, APP grows linearly with tree size, whereas the Sackin index grows quadratically  and the Total Cophenetic index cubically.

\subsection{Maximally Balanced Trees}
\label{bal}

\begin{figure*}[!htpb]
 \centering
 \includegraphics[width=0.85\textwidth]{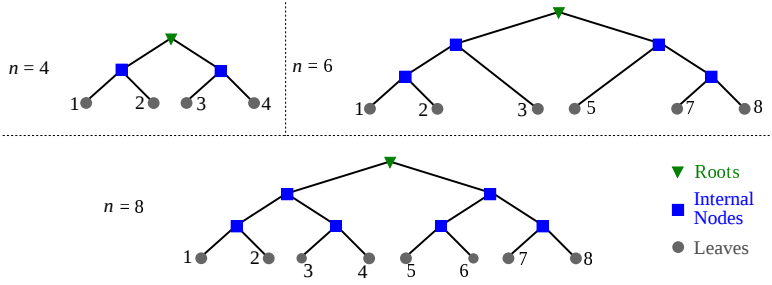}
 \caption{Maximally balanced trees for 4, 6 and 8 leaves. The green triangle represents the root node $\rho$, nodes labaled with numbers are extant species (tips) and blue squares are internal nodes.}
 \label{trees_bal}
\end{figure*}

\ \ \ \ A tree can be fully balanced only if number of leaves $n$, is such that $n=2^s$, for $s=0,1,2,...$ For $n\neq 2^s$, however, there is at least one maximally balanced tree for any value of $n$, as illustrated in Figure \ref{trees_bal}. Every fully balanced tree is also maximally balanced but the opposite is not true. Maximally balanced trees can be constructed following the algorithm in \cite{fis:2019}. Starting with a fully balanced tree of size $n'= 2^s$, leaves are removed sequentially so as to obtain a maximally balanced tree of size $n \in (2^{s-1},2^s)$. In each step of the process, the cherry with the pair of leaves $(u,v)$ (such that $d_{u,\rho}=d_{v,\rho}$, where $\rho$ is the root, and $d_{u,v}=2$) is suppressed, and its parent node becomes a leaf. Figure \ref{trees_bal} illustrates the process for $n=6$, starting with the tree of size $n=8$ and, in two steps, cherries $(3,4)$ and $(5,6)$ are deleted to form the maximally balanced tree. Exact expressions for $S_n$ and $\Phi_n$ for maximally balanced trees are given by \citep{fis:2019,mir:2013}:
\begin{equation}
 \label{Sbal}
 S_n^{(bal)} = n(\lceil\log_2 n\rceil+1)-2^{\lceil\log_2 n\rceil},
\end{equation}
\begin{equation}
 \label{Phibal}
 \Phi_n^{(bal)} = \Phi_{\lceil n/2 \rceil}^{(bal)} + \Phi_{\lfloor n/2 \rfloor}^{(bal)} + {\lceil n/2 \rceil \choose 2} + {\lfloor n/2 \rfloor \choose 2}, \quad \text{for } n>2,
\end{equation}
where $\Phi_1^{(bal)}=\Phi_2^{(bal)}=0$. Symbols $\lfloor x \rfloor$ and $\lceil x \rceil$ represent the floor and ceiling of $x$.  The APP index can now be computed from eqs. (\ref{wSPhi}), (\ref{Sbal}) and (\ref{Phibal}).\\

For the fully balanced case ($n=2^s$) we can also write closed formulas for all these expressions. For simplicity we write $S_{2^s}^{(bal)}=S_s$ and $\Phi_{2^s}^{(bal)}=\Phi_s$:
\begin{equation}
 S_s  = s2^s,
\end{equation}
\begin{equation}
 \label{Phibalr}
 \Phi_s = 2\Phi_{s-1} + 2^{s-1}(2^{s-1} - 1), \quad \text{for } s>1,
\end{equation}
with $\Phi_{s=0}=\Phi_{s=1}=0$.  Solving this recursive relation we also obtain 
\begin{equation}
 \label{Phibals}
 \Phi_s = 2^{s-1}(2^s-s-1).
\end{equation}
We finally obtain the expression for the APP index. For simplicity we write $\bar{d}_{2^s}^{(bal)}=\bar{d}_s$.
\begin{eqnarray}
 \label{wbals}
 \bar{d}_s &=& \frac{S_s}{2^{s-1}}-\frac{\Phi_s}{2^{s-2}(2^s-1)}, \nonumber \\
 & = & \frac{s2^s}{2^{s-1}}-\frac{2^{s-1}(2^s-s-1)}{2^{s-2}(2^s-1)}, \nonumber \\
 & = & \frac{2}{2^s-1}[1+2^s(s-1)], \quad \text{for } s>0,
\end{eqnarray}
with $\bar{d}_{s=0}=0$. Using eq. (\ref{wbals}) we can make approximations for maximally balanced trees which do not have $n=2^s$, but $n\simeq2^s$:
\begin{equation}
 \label{wbaln}
 \bar{d}_n^{(bal)} \simeq \frac{2}{n-1}[1+n(\log_2 n-1)], \quad \text{for } n>1,
\end{equation}
where $\bar{d}_1^{(bal)}=0$. In Table \ref{tab} we show the accuracy of this approximation for several values of $n$. For large $n$ the APP index for maximally balanced trees grows logarithmically with tree size  ($\bar{d}_n^{(bal)} \sim \log n$), which is slower than the Sackin ($S_n^{(bal)} \sim n \log n$) and Total Cophenetic indices ($\Phi_n^{(bal)} \sim n^2$). 
\begin{table}
 \caption{Comparison between exact ($=$) and approximated ($\simeq$) formulas to APP index for maximally balanced trees. The values highlighted(*) are for fully balanced trees, where the approximation becomes exact.}
 \label{tab}
 \begin{tabular}{lllllllllll}
  \hline\noalign{\smallskip}
  $n$ & 2* & 3 & 4* & 5 & 6 & 7 & 8* & 9 \\
  \noalign{\smallskip}\hline\noalign{\smallskip}
  = & 2 & 2.66667 & 3.33333 & 3.8 & 4.26667 & 4.57143 & 4.85714 & 5.16667 \\ 
  $\simeq$ & 2 & 2.75489 & 3.33333 & 3.80482 & 4.20391 & 4.55049 & 4.85714 & 5.13233 \\ 
  \noalign{\smallskip}\hline
 \end{tabular}
\end{table}\\

\begin{figure*}[!htpb]
 \centering
 \includegraphics[width=0.75\textwidth]{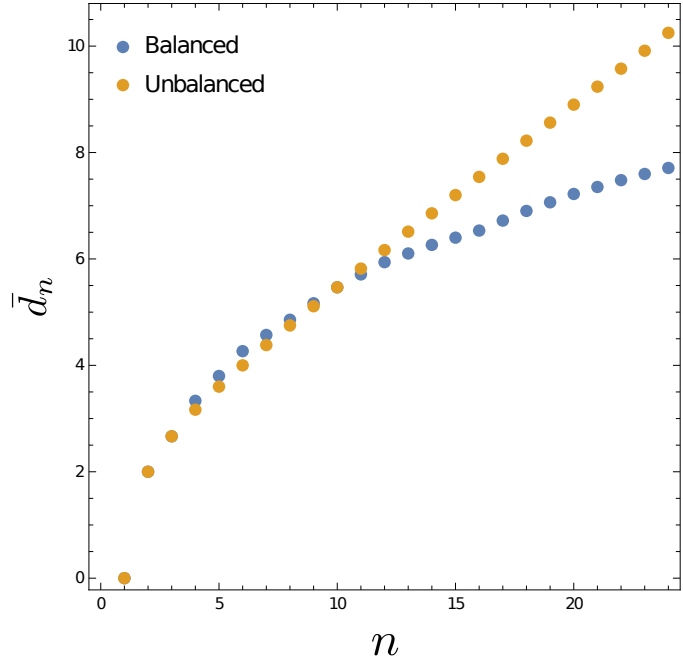}
 \caption{APP index for fully unbalanced and maximally balanced trees for some number of leaves -- $n$. For $1\leq n\leq3$, the index coincides for the two cases since the trees are the same. For $4\leq n\leq9$, the index for balanced trees is larger than for unbalanced. For $n=10$, the values  coincide again, while for $n>10$, the index for unbalanced trees is larger than for balanced trees.}
 \label{wcases}
\end{figure*}

For the Sackin and Total Cophenetic indices, the maximum and minimum values are reached for fully unbalanced and maximally balanced trees with size $n$, respectively \citep{fis:2019,mir:2013}. Figure \ref{wcases} shows that for $n>10$ $\bar{d}_n$ is larger for fully unbalanced trees than for maximally balanced ones. This indicates that APP index should be a good metric for measuring balance only for large $n$. In the next section, we present exact formulas for $\bar{d}_n$ under the Yule Model.

\section{APP Index under the Yule Model}
\label{yule}

\subsection{Expected Value}
\label{expec}

\ \ \ \ The Yule Model is a simple algorithm to generate stochastic binary trees \citep{har:1971,kir:1993,ste:2001,car:2013}. The Yule algorithm starts with a cherry (a root and two leaves) and at each step of the process one tip is selected with uniform probability and is replaced by a cherry with two new tips. In this process different trees with the same number of tips are generated with different frequencies. The Yule model describes ensembles of random trees, with no specific elements affecting the evolutionary process. The expected value and variance for the Sackin \citep{kir:1993,ste:2001} and Total Cophenetic indices \citep{car:2013} can be computed directly for such ensembles.\\

\begin{figure*}[!htpb]
 \centering
 \includegraphics[width=0.47\textwidth]{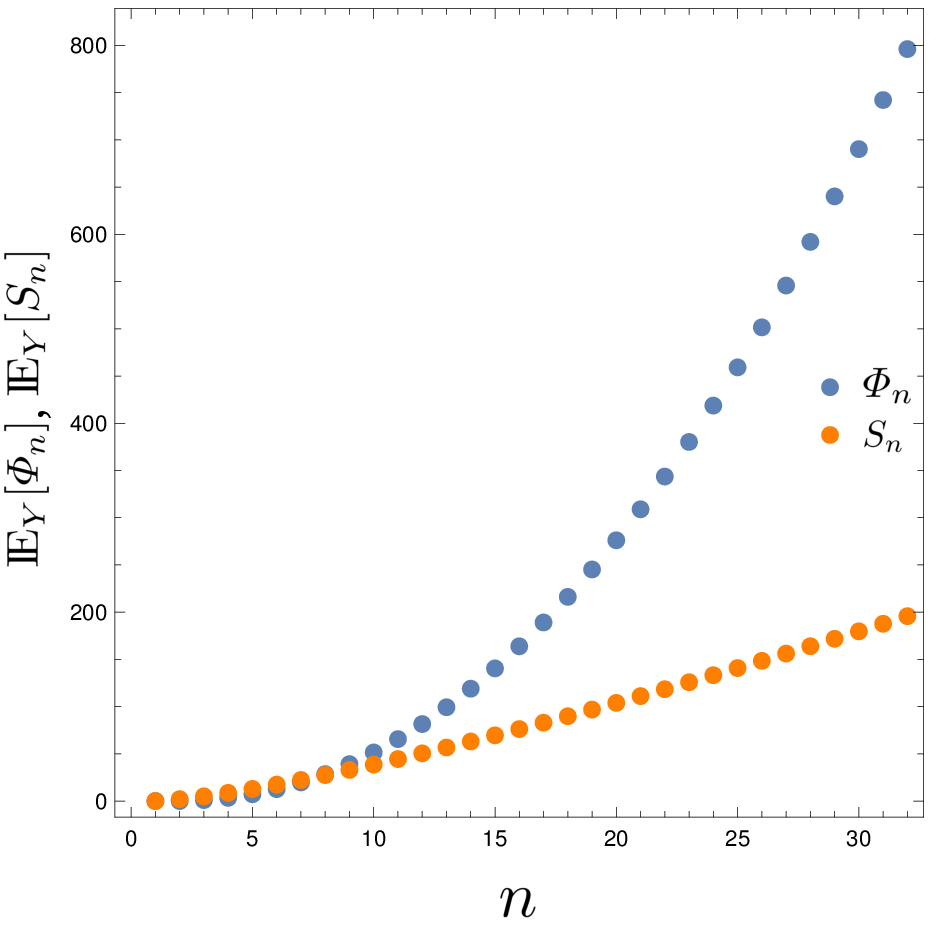} \qquad \includegraphics[width=0.46\textwidth]{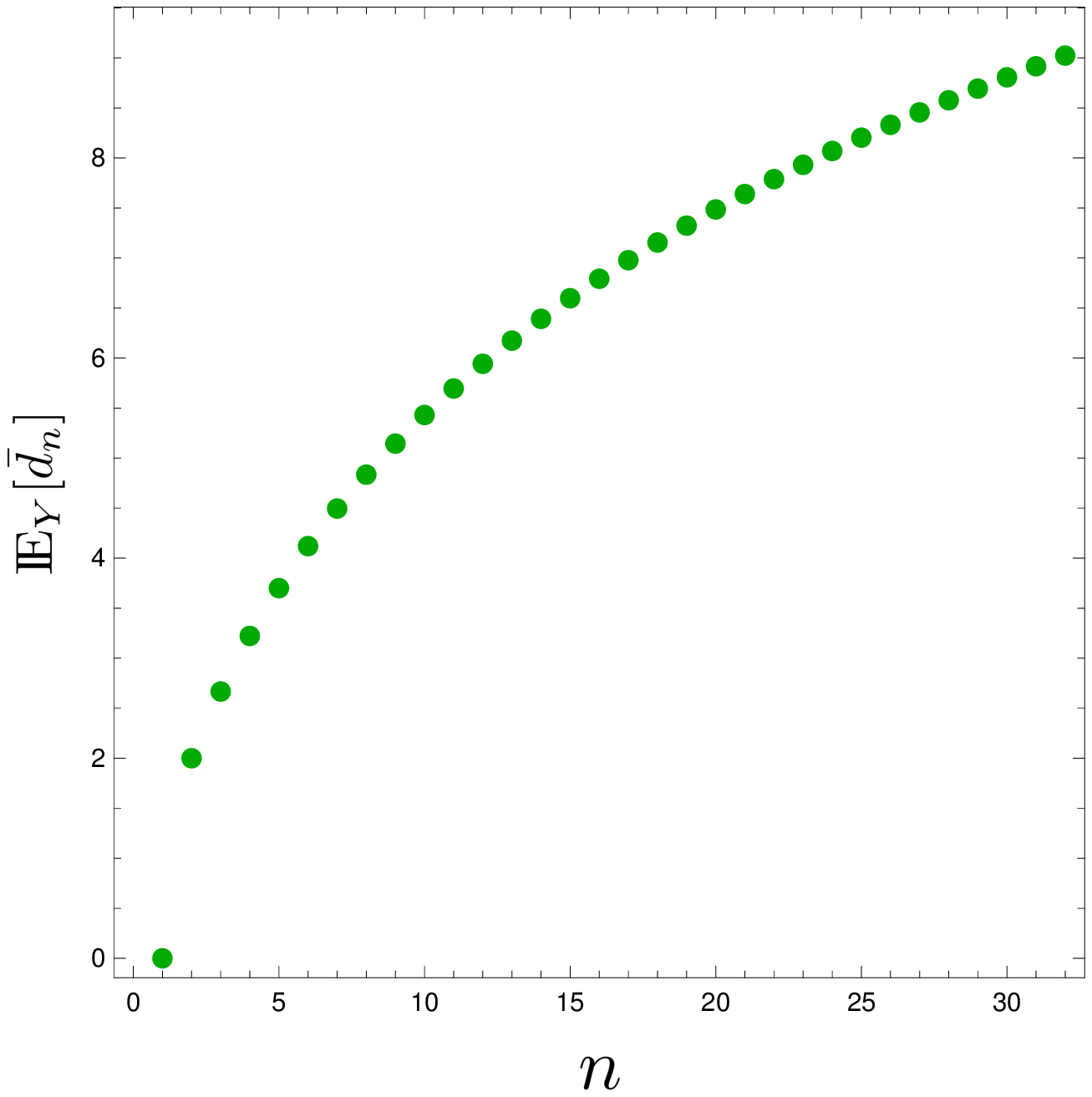}
 \caption{Left Panel: expected value under the Yule Model of Sackin and Total Cophenetic indices as a function of $n$. Right Panel: same for APP index.}
 \label{EwSPhi}
\end{figure*}

Let $\bar{d}_n$ be a random variable associated with a tree of size $n$ created under the Yule model and whose value is the APP index of that tree. To calculate its expected value $\E_Y[\bar{d}_n]$ we will resort to expressions demonstrated previously in \cite{car:2013} for the expected values of the Sackin and Total Cophenetic indices:
\begin{equation}
 \label{ES}
 \E_Y[S_n]=2n \left [ H_n^{(1)}-1 \right ],
\end{equation}
\begin{equation}
 \label{EPhi}
 \E_Y[\Phi_n]=n(n-1)-2n \left [ H_n^{(1)}-1 \right ],
\end{equation}
where $H_n^{(r)}=\sum_{i=1}^n 1/i^r$ is the $n$th Harmonic Number of order $r$. Substituting the formulas above in eq. (\ref{wSPhi}) we obtain the expected value of $\bar{d}_n$:
\begin{eqnarray}
 \label{Ew}
 \E_Y[\bar{d}_n] &=& \frac{2}{n}\E_Y[S_n] - \frac{4}{n(n-1)}\E_Y[\Phi_n], \nonumber \\
 & = & \frac{2}{n}\cdot2n \left [H_n^{(1)}-1 \right ] - \frac{4}{n(n-1)}\cdot \left \{n(n-1)-2n \left [ H_n^{(1)}-1 \right ] \right \}, \nonumber \\
 & = & 4 \left \{ \left [ H_n^{(1)}-1 \right ] \left ( \frac{n+1}{n-1} \right ) - 1 \right \}, \quad \text{for } n>1,
\end{eqnarray}
where $\E_Y[\bar{d}_1]=0$. For large $n$ the asymptotic behavior of the expected value of APP index under the Yule model has positive growth with tree size -- n ( $\E_Y[\bar{d}_n]  \sim \log n$). The growth with tree size is slower than the expected values for Sackin ($\E_Y[S_n]  \sim n \log n$) and for Total Cophenetic indices ($\E_Y[\Phi_n]\sim n^2$). Figure \ref{EwSPhi} shows the behavior of these expected values as a function of $n$.

\begin{figure*}[!htpb]
 \centering
 \includegraphics[width=0.43\textwidth]{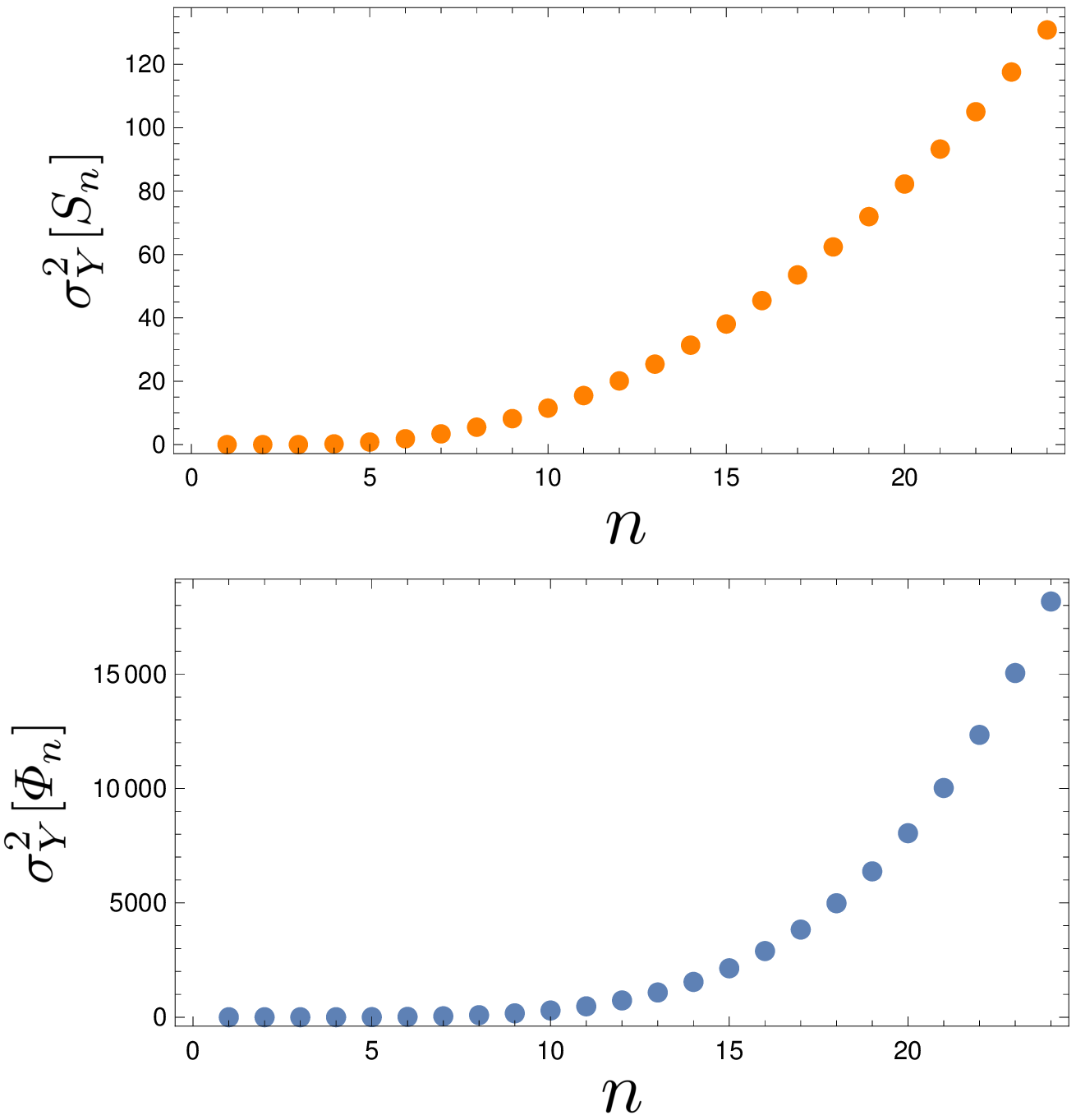} \qquad \includegraphics[width=0.47\textwidth]{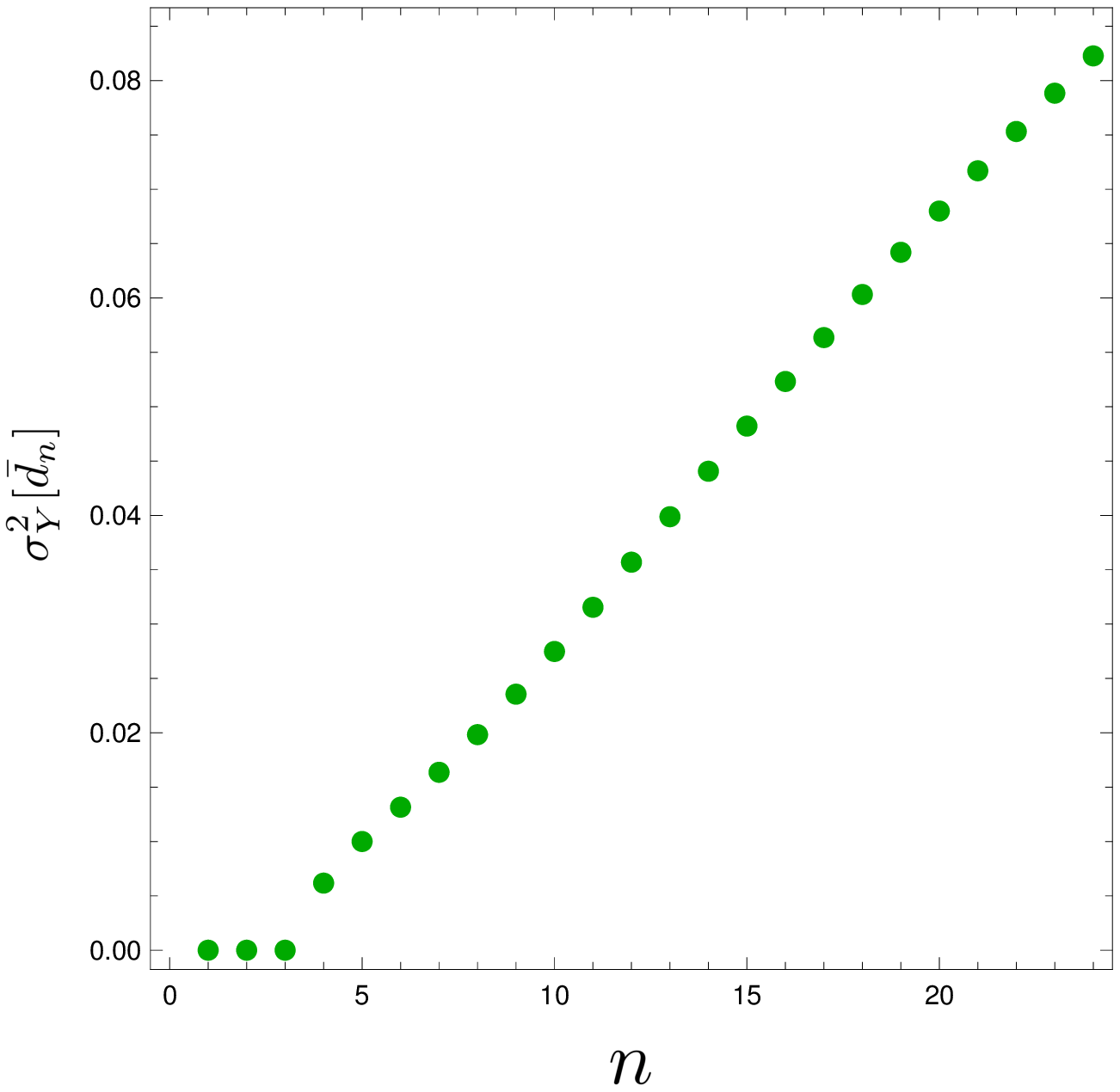}
 \caption{Left Upper Panel: variance of the Sackin index under the Yule Model. Left Bottom Panel: same for the Total Cophenetic index. Right Panel: same for APP index.}
 \label{varfig}
\end{figure*}

\begin{figure*}[!htpb]
 \centering
 \includegraphics[width=0.43\textwidth]{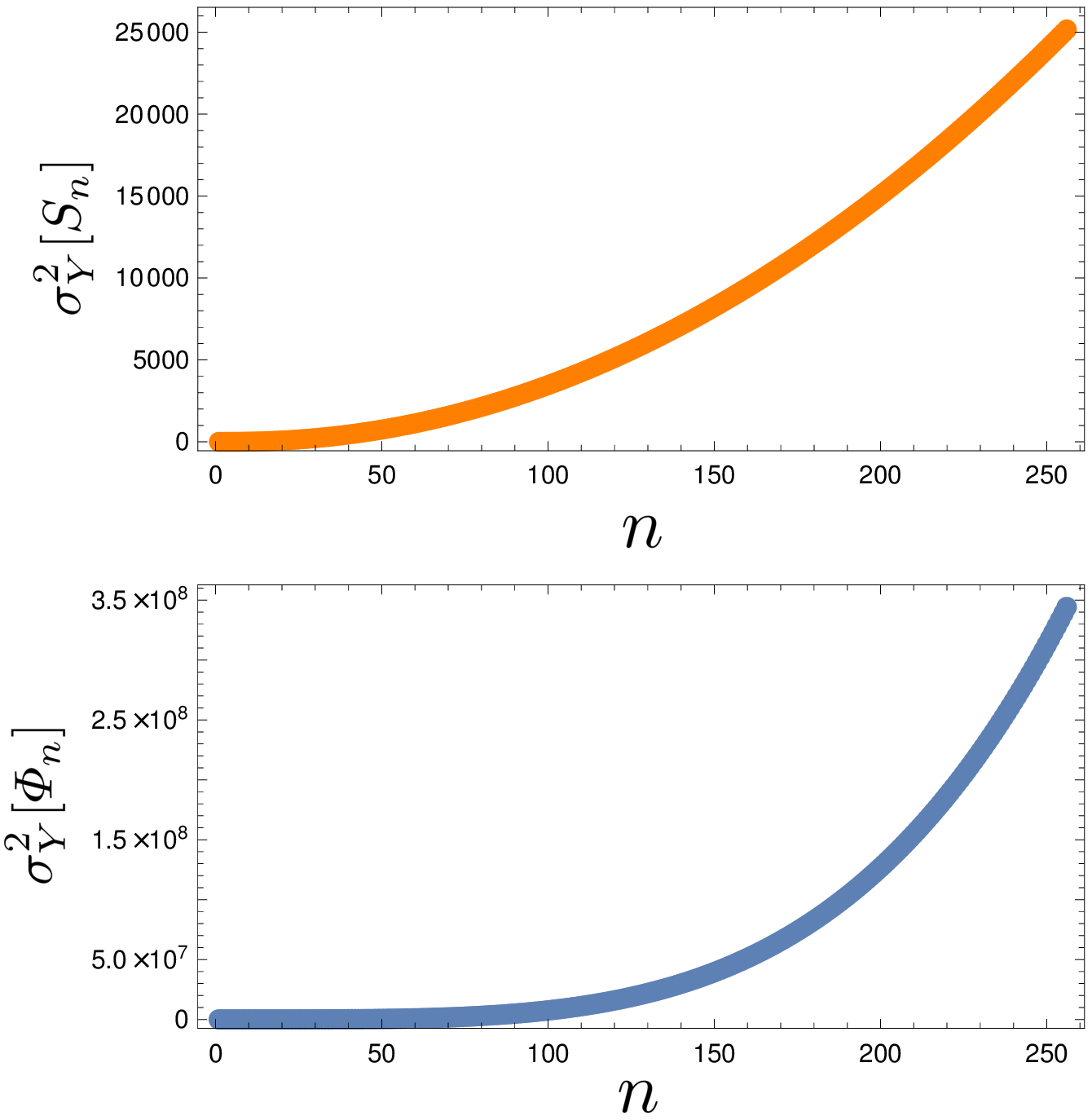} \qquad \includegraphics[width=0.47\textwidth]{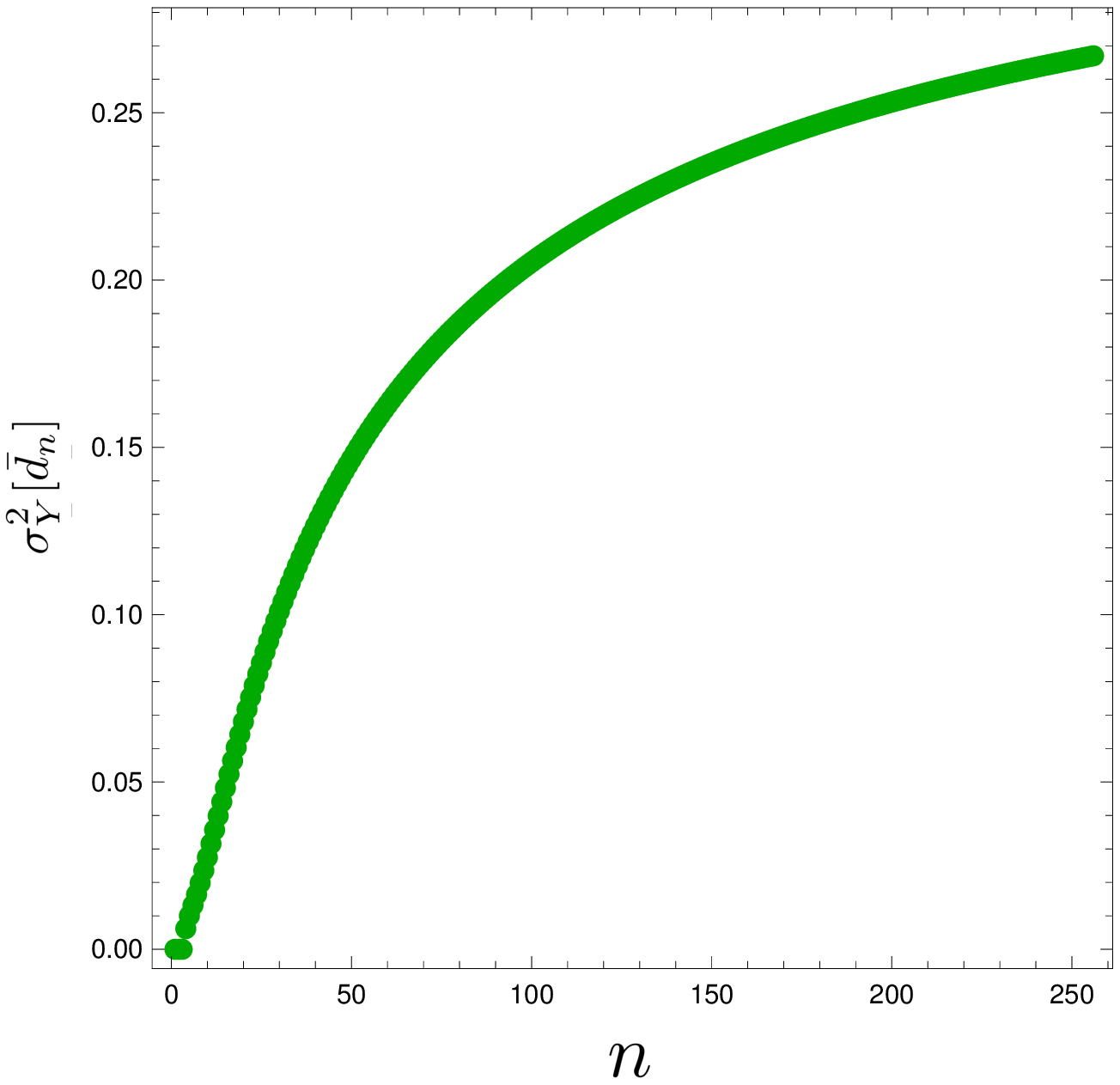}
 \caption{Left Upper Panel: variance of the Sackin Index under the Yule Model. Left Bottom Panel: same for the Total Cophenetic index. Right Panel: same for APP index.  For large $n$, $\sigma_Y^2[d_n]$ converges to an asymptotic value.}
 \label{varasym}
\end{figure*}

\begin{figure*}[!htpb]
 \centering
 \includegraphics[width=0.44\textwidth]{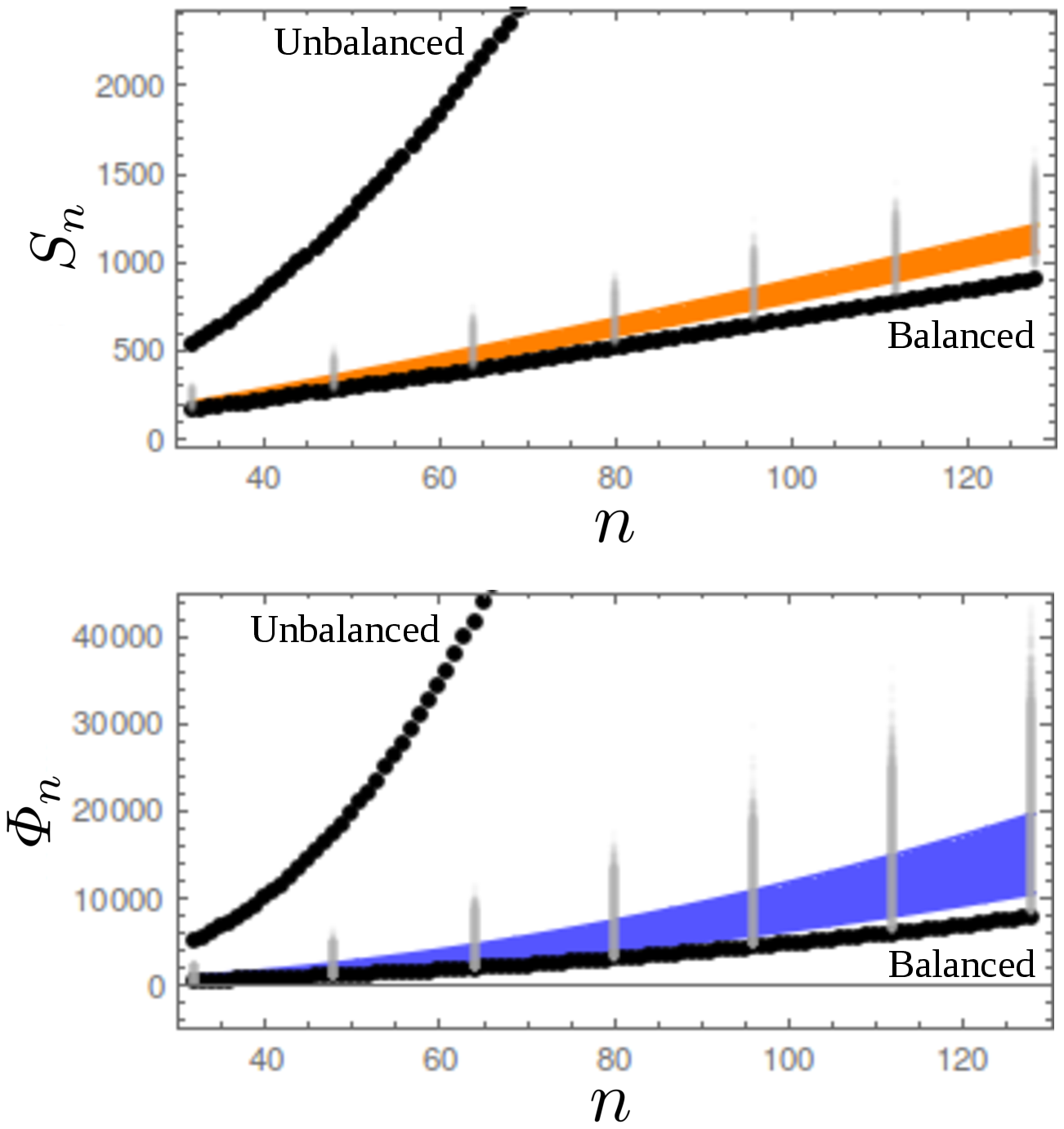} \qquad \includegraphics[width=0.47\textwidth]{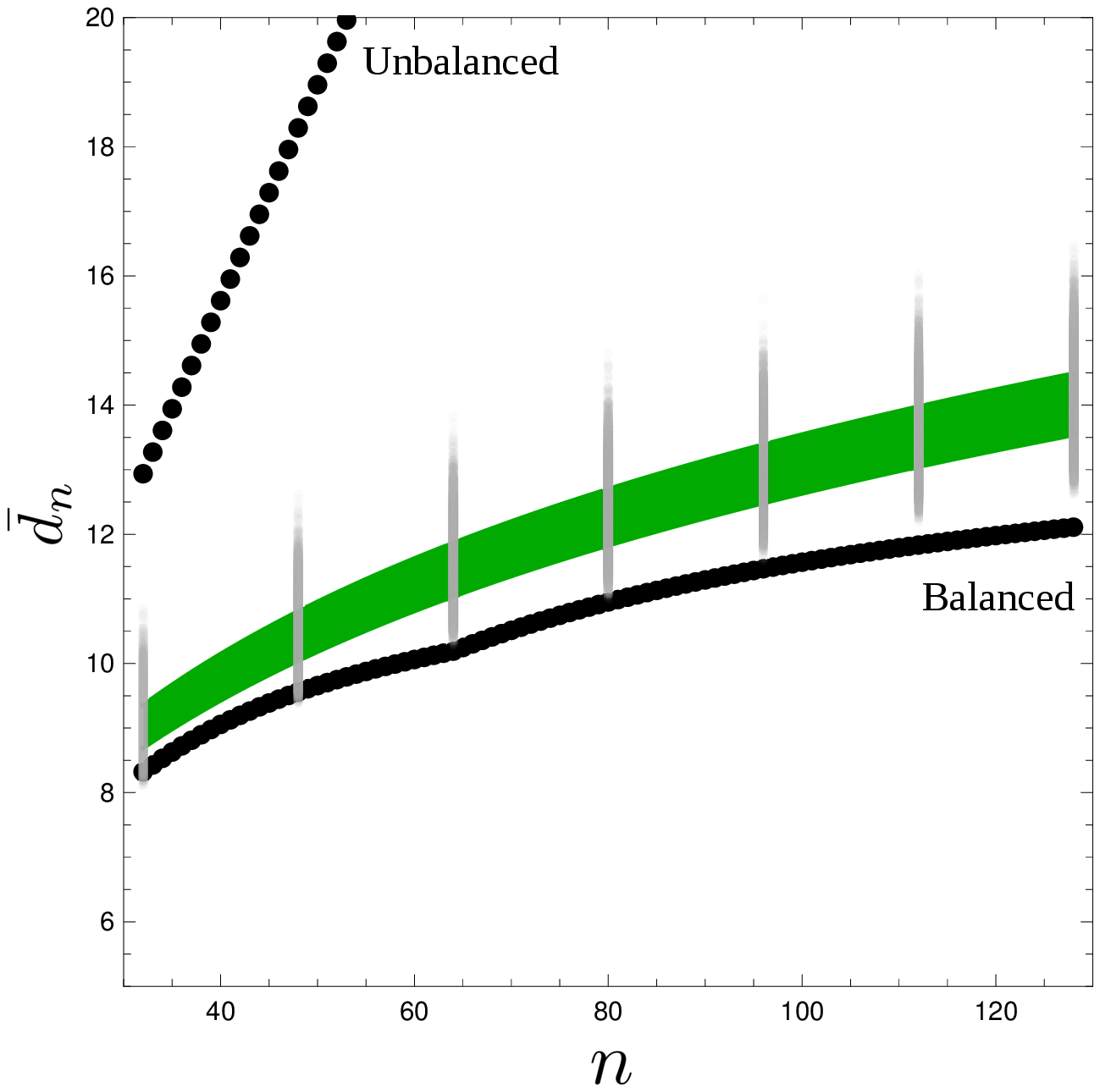}
 \caption{Black dots: values of indices for the fully unbalanced and maximally balanced trees. Gray dots: values for an ensemble of 50,000 trees obtained from the Yule Model. Left Upper Panel: Sackin Index. Left Bottom Panel: Total Cophenetic index. Right Panel: APP index. Orange, blue and green regions represent the interval of one standard deviation around the mean for the Sackin, Total Cophenetic and APP indeces respectively.}
 \label{ensemble}
\end{figure*}

\subsection{Variance}
\label{var}

\ \ \ \ Let $I_n$ be a balance index for a phylogenetic tree with $n$ leaves and $\E_Y[I_n]$ and $\sigma^2_Y[I_n]$ the expected value and variance of this index under the Yule Model respectively. In order to compare trees with different number of leaves it is common to define a normalized index \citep{blu:2006} as: 
\begin{equation}
 \label{Inorm}
 I_n^{(norm)} = \frac{I_n-\E_Y[I_n]}{\sqrt{\sigma^2_Y[I_n]}}.
\end{equation}
Clearly $\E_Y[I_n^{(morm)}]=0$ and $ \sigma_Y^2[I_n^{(norm)}] = 1$. To construct a normalized index for $\bar{d}_n$ we must calculate its variance under the Yule Model, $\sigma_Y^2[\bar{d}_n]$. The following property of variance will be useful in our development:
\begin{equation}
 \label{varprop}
 \sigma^2[AX+BY] = A^2\sigma^2[X]+B^2\sigma^2[Y] + 2AB\Cov[X,Y],
\end{equation}
where $X$ and $Y$ are random variables, $A$, and $B$ are constants and $\Cov[X,Y]$ is the covariance of $X$ and $Y$. Applying this property in eq. (\ref{wSPhi}) we get
\begin{equation}
 \label{varw1}
 \sigma_Y^2[\bar{d}_n] = \frac{4}{n^2} \sigma_Y^2[S_n] + \frac{16}{n^2(n-1)^2} \sigma_Y^2[\Phi_n] - \frac{16}{n^2(n-1)} \Cov_Y[S_n,\Phi_n].
\end{equation}
The quantities $\sigma_Y^2[S_n]$, $\sigma_Y^2[\Phi_n]$ and $\Cov_Y[S_n,\Phi_n]$ were elegantly calculated using the fact that these indices satisfy recursive relations by \cite{car:2013}. The final expressions are:
\begin{equation}
 \label{varS}
 \sigma_Y^2[S_n] = n^2 \left [ 7 - 4H_n^{(2)} \right ] - n \left [ 1 + 2H_n^{(1)} \right ],
\end{equation}
\begin{equation}
 \label{varPhi}
 \sigma_Y^2[\Phi_n] = \frac{n}{12}(n^3-10n^2+131n-2) - 2n \left [ 2n H_n^{(2)} + 3 H_n^{(1)} \right ],
\end{equation}
\begin{equation}
 \label{CovSPhi}
 \Cov[S_n,\Phi_n] = \frac{n}{6}(n^2-51n+2) + 4n \left [ n H_n^{(2)} + H_n^{(1)} \right ].
\end{equation}
Finally, the variance under the Yule Model of the APP index ($\bar{d}_n$) is
\begin{eqnarray}
 \label{varw}
 \sigma_Y^2[\bar{d}_n] = \frac{4}{3n(n-1)^2} \Bigl [ 20n^3 + 49n^2 + 52n - 1 - \nonumber\\
 6(n+1)(n+5)H_n^{(1)} - 12n(n+1)^2H_n^{(2)} \Bigr ].
\end{eqnarray}
Figure \ref{varfig} shows the behavior of $\sigma_Y^2[\bar{d}_n]$ in comparison with the variance of Sackin and the Total Cophenetic indices. For $n\gg1$, this variance converges to an asymptotic value:
\begin{equation}
 \label{varwass}
 \sigma_Y^2[\bar{d}_n] \simeq \frac{4}{3n^3} \left [ 20n^3 - 12n^3 \frac{\pi^2}{6} \right ] = \frac{8}{3} (10-\pi^2)  \simeq 0.3477 ,
\end{equation}
where we used $H_n^{(2)} \simeq \zeta(2) = \pi^2/6$ for $n\gg1$ and $\zeta(r)$ is the Riemman Zeta Function. This asymptotic behavior is the main difference between the APP index and the other two metrics. For large trees the variance of Sackin and Total Cophenetic indices are
\begin{equation}
 \label{varSass}
\sigma_Y^2[S_n] \simeq n^2 \left [ 7 - 4\frac{\pi^2}{6} \right ] \simeq 0.4203 n^2
\end{equation}
and
\begin{equation}
 \label{varPhiass}
\sigma_Y^2[\Phi_n] \simeq \frac{n^4}{12} \simeq 0.0833 n^4.
\end{equation}
Figure \ref{varasym} shows the variances for $1 \leqslant n \leqslant 256$.  Convergence of the APP index variance is very slow;  for $n=6000$, for instance, $\sigma_Y^2[\bar{d}_n] \simeq 0.3402$. Figure \ref{ensemble} shows results for an ensemble of 50,000 trees generated with the Yule Model for several values of $n$ where the Sackin, Total Cophenetic, and APP indices were computed. In the left panels (Sackin -- $S_n$ and Total Cophenetic -- $\Phi_n$), the dispersion of the distribution for each $n$ increases with $n$, more prominently for  $\Phi_n$ in the bottom panel. In contrast, on the right panel (APP index $\bar{d}_n$), the dispersion is nearly constant, increasing on slightly with $n$.\\

\begin{figure*}[!htpb]
 \centering
 \includegraphics[width=0.42\textwidth]{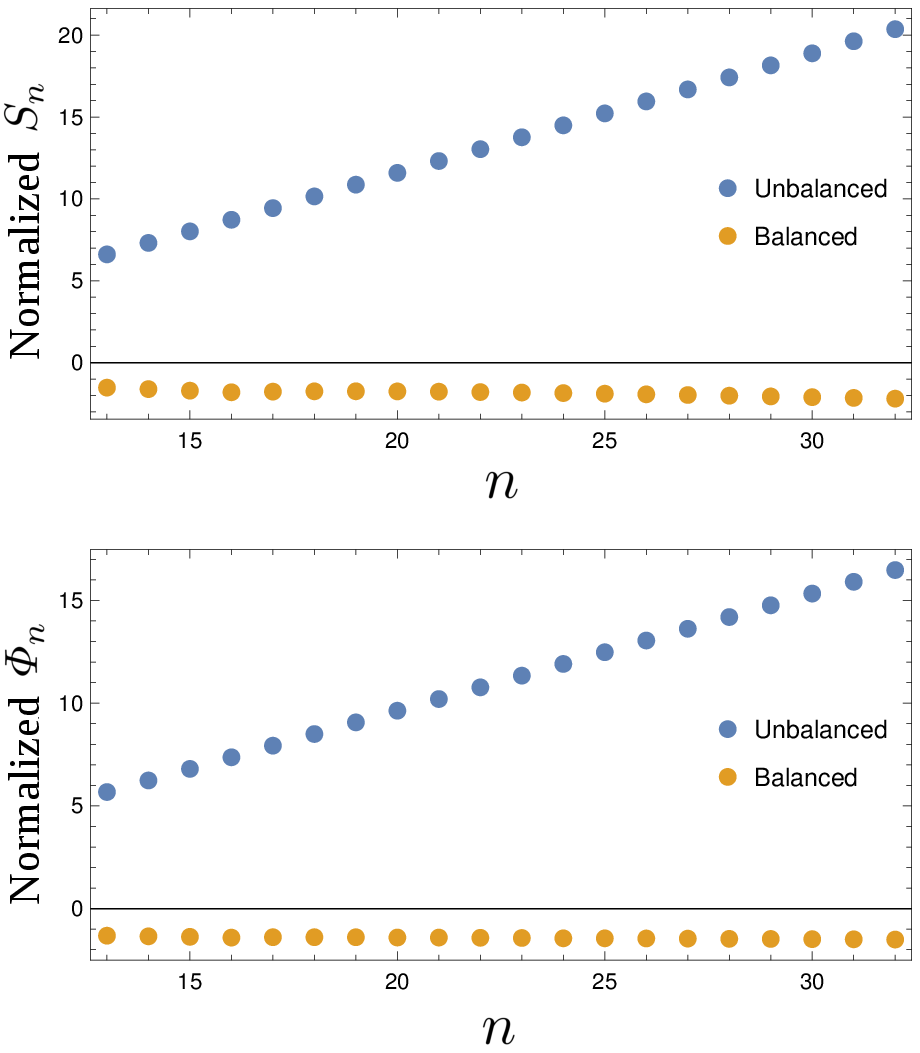} \qquad \includegraphics[width=0.48\textwidth]{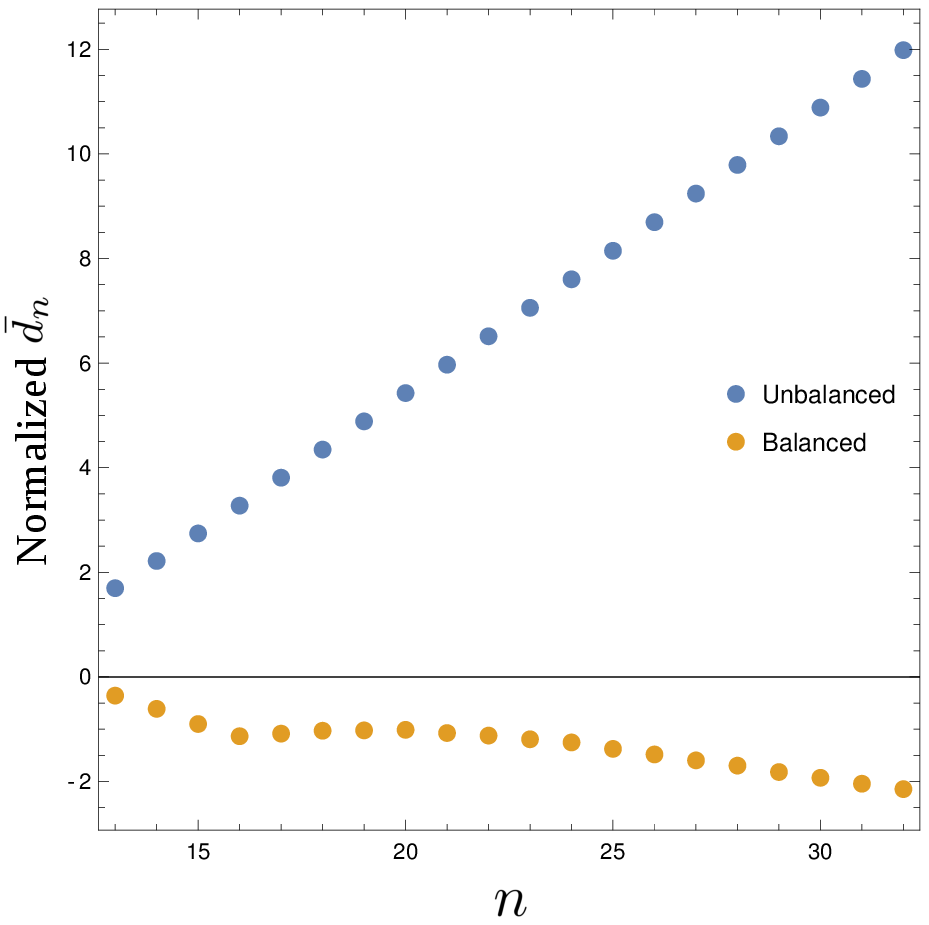}
 \caption{Left Upper Panel: Maximum (unbalanced trees) and minimum (balanced trees) for The normalized Sackin index. Left Bottom Panel: same for the normalized Total Cophenetic index. Right Panel: same for the Normalized APP index. For all indices, the unbalanced and balanced cases depart from zero in very similar ways.}
 \label{normcasesfig}
\end{figure*}

Although there is a remarkable difference between the variance of the different metrics, their normalized versions (according to eq. (\ref{Inorm})) erase all relevant distinctions. For $n \geqslant 10$, we plot the maximum (unbalanced trees) and minimum (balanced trees) value of each normalized index. As expected, the spread between the maximum and minimum values is similar for the three indices, Figure \ref{normcasesfig}.

\begin{figure*}[!htpb]
 \centering
 \includegraphics[width=0.75\textwidth]{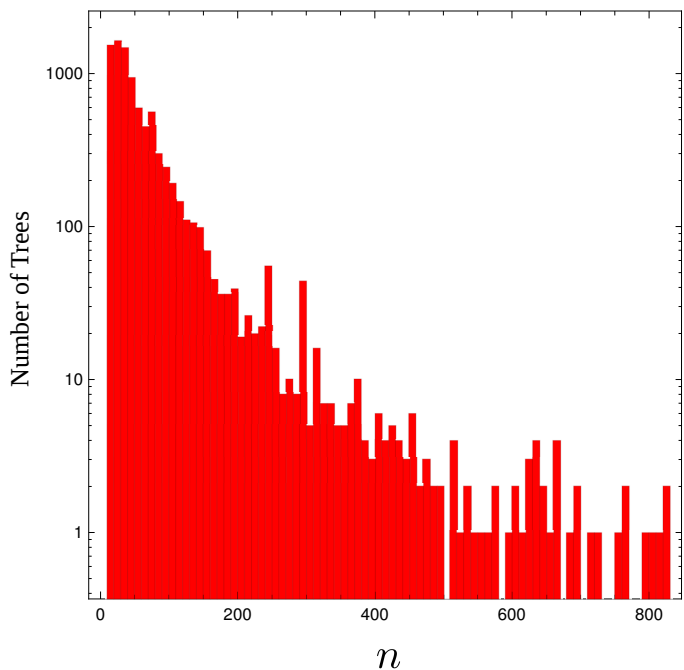}
 \caption{Distribution of the number of tips (leaves) in the set of real phylogenies used for comparison. After filtering the original set, regarding only species trees with $n \in [10,900]$, we analyzed the three indices in a set of 8999 empiric trees.}
 \label{fig:disttips}
\end{figure*}

\begin{figure*}[!htpb]
 \centering
 \includegraphics[width=0.43\textwidth]{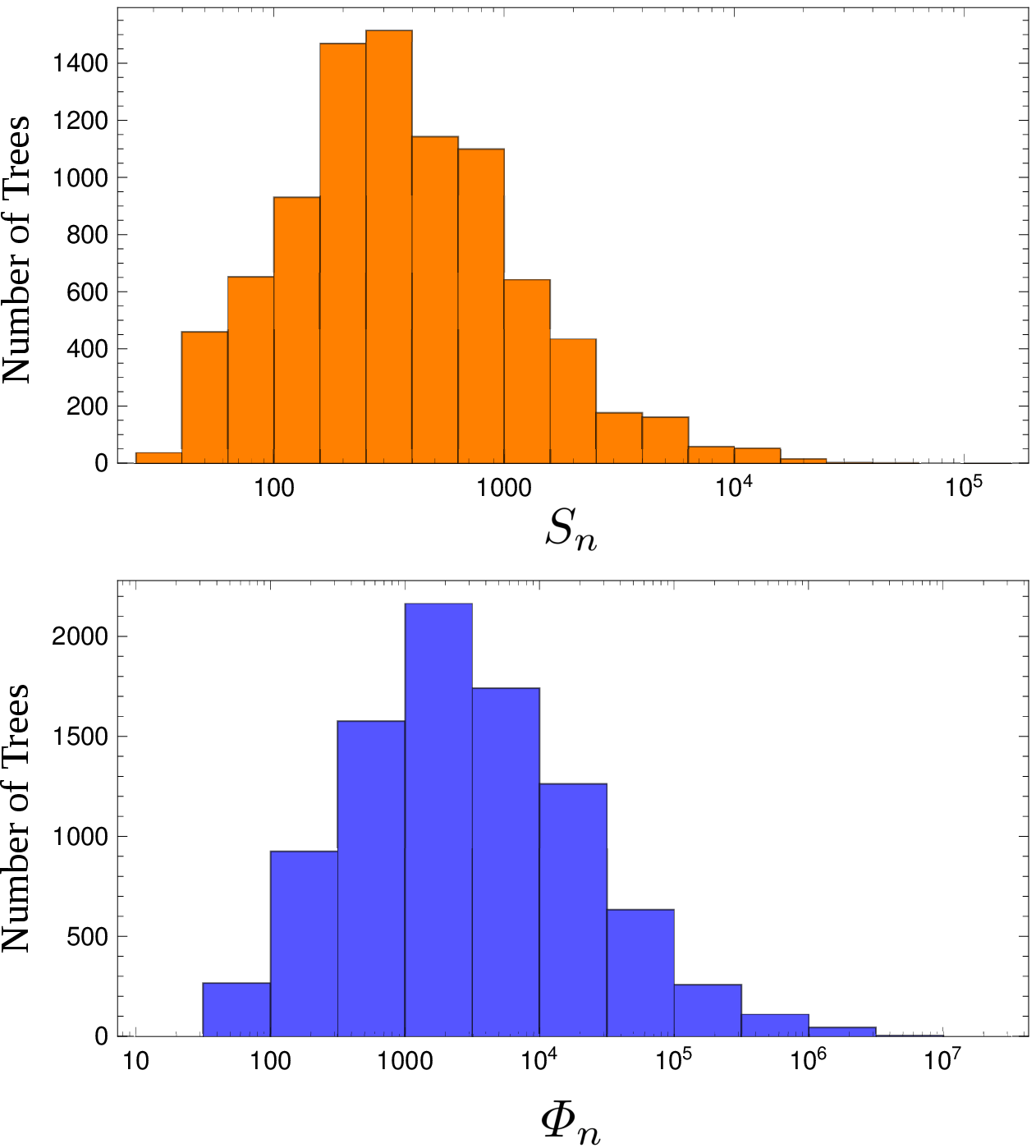} \qquad \includegraphics[width=0.48\textwidth]{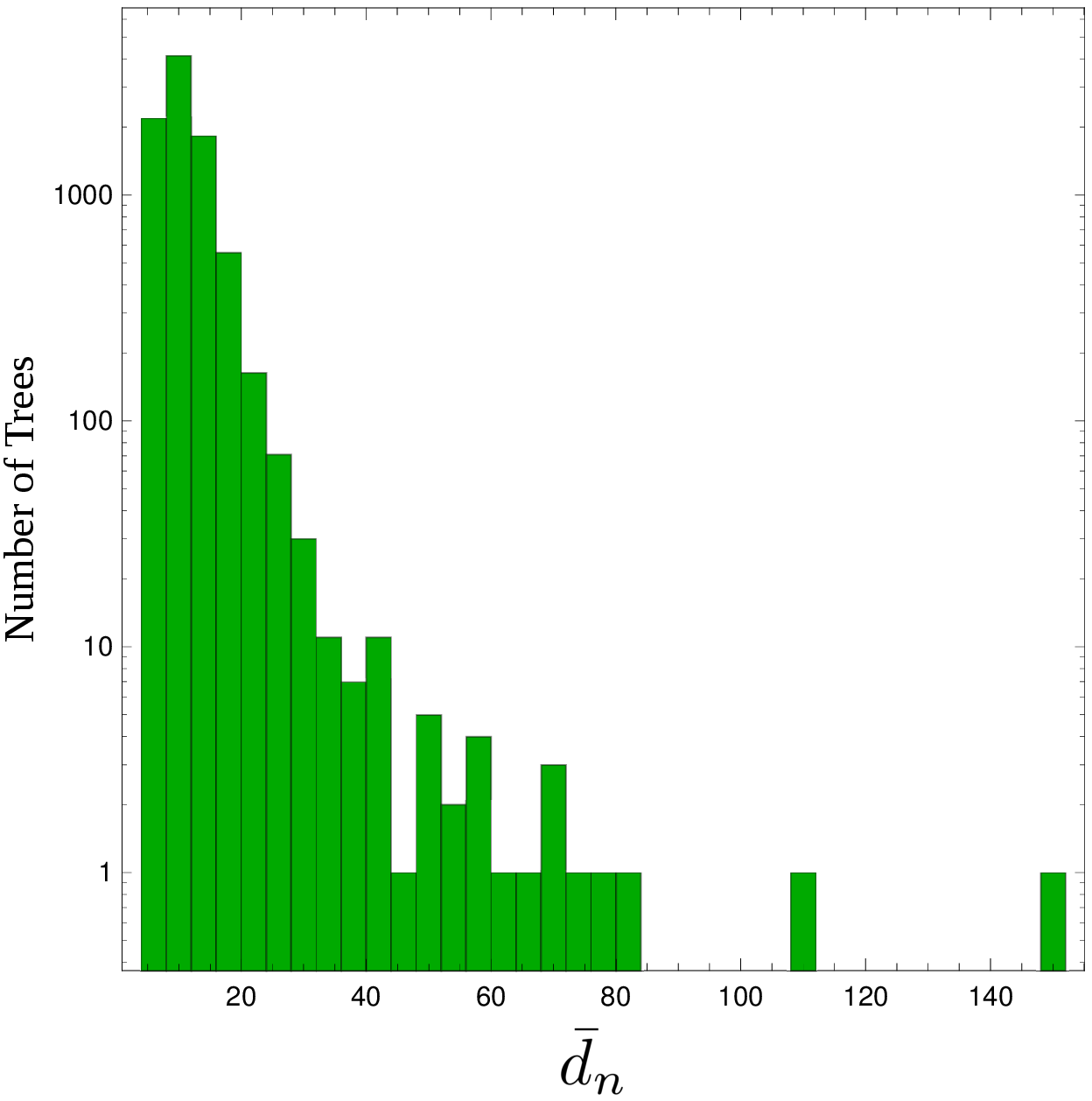}
 \caption{Left Upper Panel: distribution of trees with Sackin Index $S_n$. Left Bottom Panel: same for the Total Cophenetic index $\Phi_n$. Right Panel: same for the APP index $\bar{d}_n$. Our proposal shows a larger concentration of trees in low values, while the previous indices present a larger concentration in intermediate values.}
 \label{fig:distmetrics}
\end{figure*}

\begin{figure*}[!htpb]
 \centering
 \includegraphics[width=0.43\textwidth]{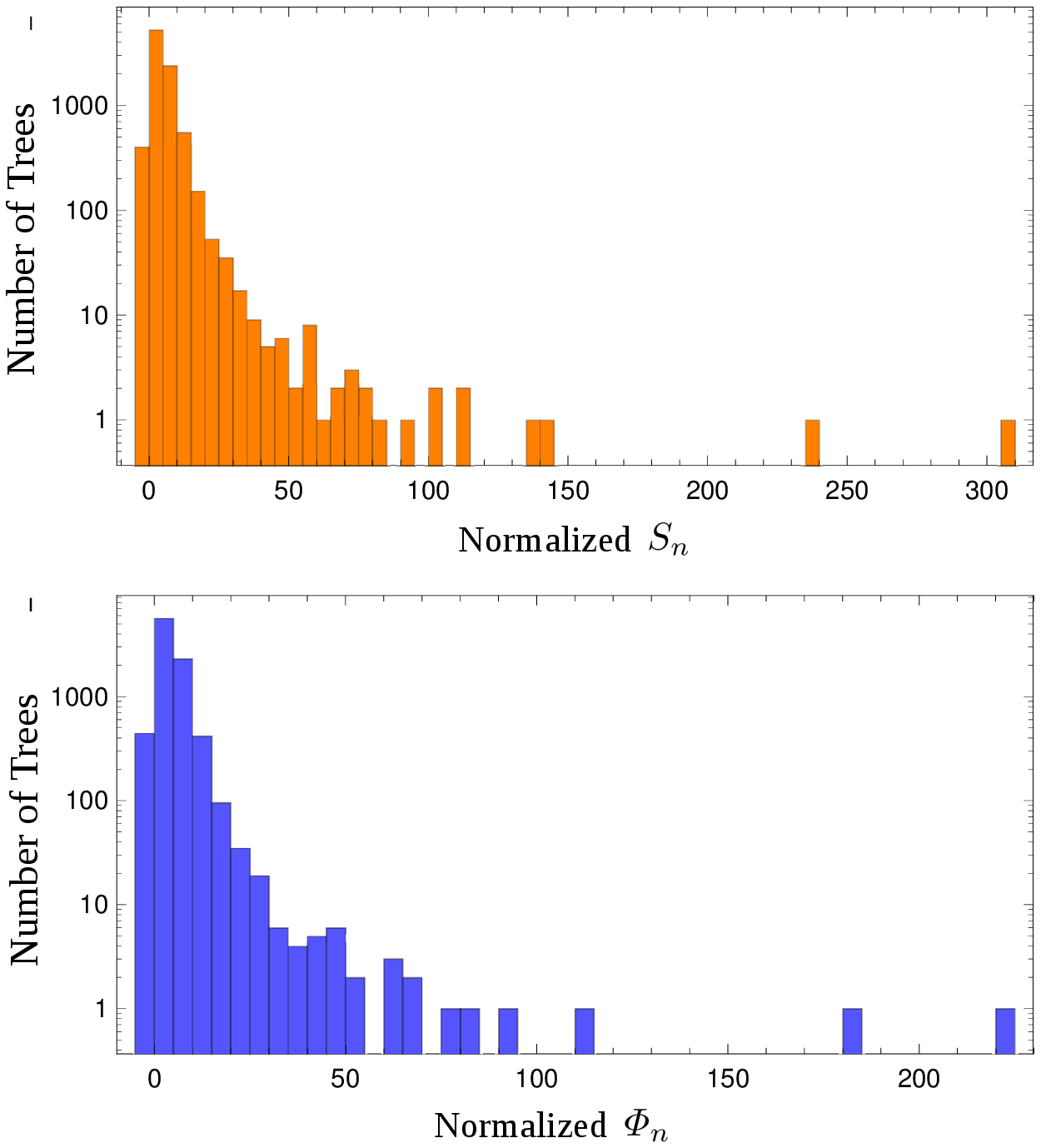} \qquad \includegraphics[width=0.48\textwidth]{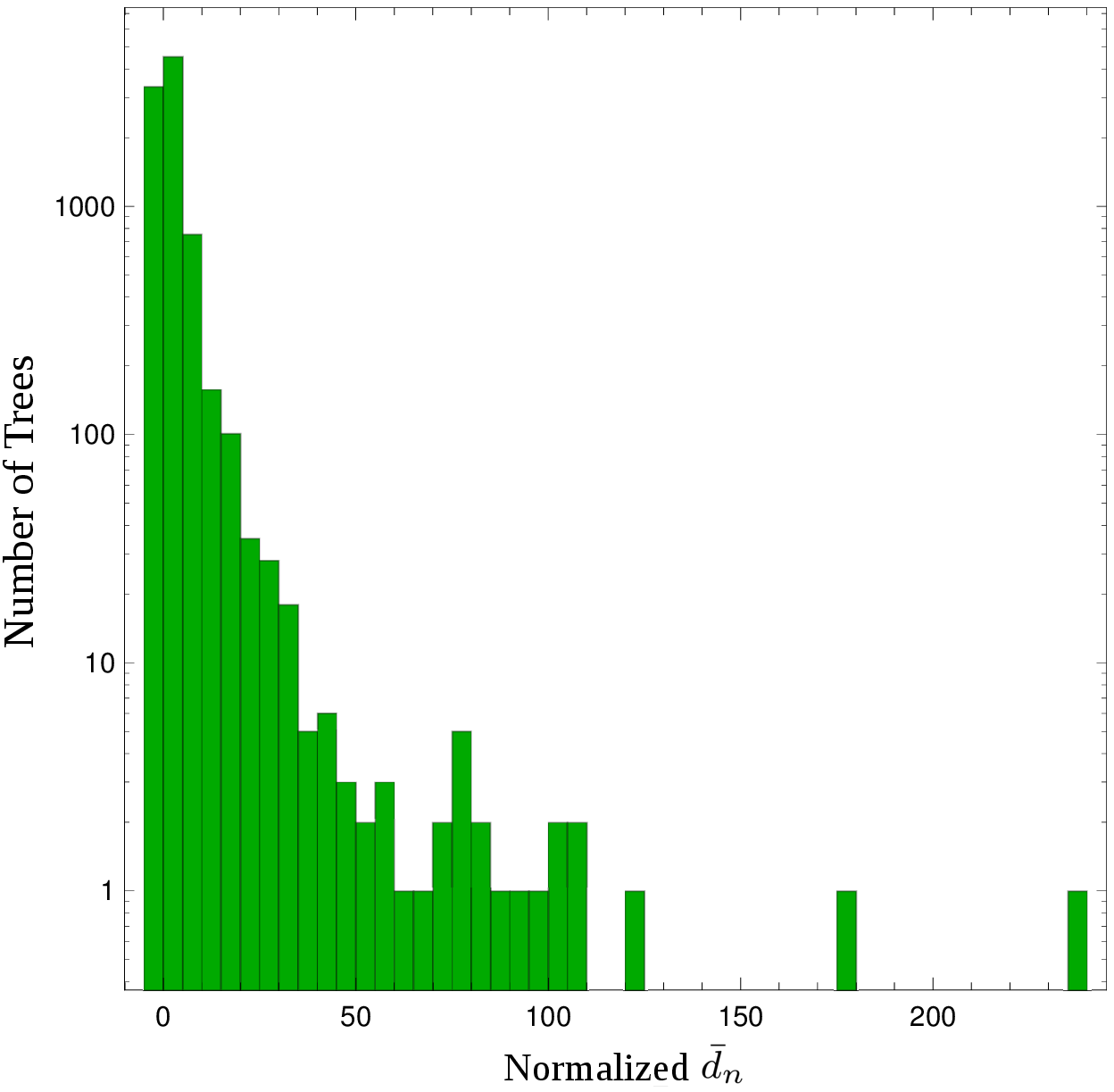}
 \caption{Left Upper Panel: distribution of trees with normalized Sackin Index $S_n^{(norm)}$. Left Bottom Panel: same for the Total Cophenetic index $\Phi_n^{(norm)}$. Right Panel: same for the APP index $\bar{d}_n^{(norm)}$. The normalization procedure described in eq. (\ref{Inorm}) tends to equalize the distribution of the three indices.}
 \label{fig:distmetricsnorm}
\end{figure*}

\section{Empiric Trees}
\label{real}

\ \ \ \ In order to understand how the distribution of the different balance metrics represents the distribution of real phylogenetic trees, we analyzed empiric phylogenetic trees available at the TreeBASE \citep{pie:2000,vos:2012}. We generated a subset among the available trees using three filters: in the field \textit{kind} we used ``Species Tree'' (no ``Gene Tree'' and ``Barcode Tree'' were included), and in the field \textit{type} we used ``Single'' (no consensus tree were included). With this procedure, we end up with 9805 phylogenetic trees. Among these trees, we created a subset based on the number of tips, by defining the \textit{ntaxa} field with trees above ten species and below 900 species. After this filtering, we ended up with 9307 trees. Of this subset, only 8999 trees were used, because 308 trees had issues when reading the files from the database for calculating the metrics. We used R \citep{R:2017} for reading and calculating the balance metrics for the empiric trees. We show the number of tips distribution of the empiric trees in the Figure \ref{fig:disttips}. Figures \ref{fig:distmetrics} and \ref{fig:distmetricsnorm} show the distribution of the balance indices for the empiric trees. Our index proposal (APP) shows a larger concentration of trees in low values of the index, while the Sackin and Total Cophenetic indices present a larger concentration in intermediate values. After the normalization procedure described in eq. (\ref{Inorm}), the three normalized indices investigated tends to show very similar distributions.
\\

\begin{figure*}[!htpb]
 \centering
 \includegraphics[width=0.43\textwidth]{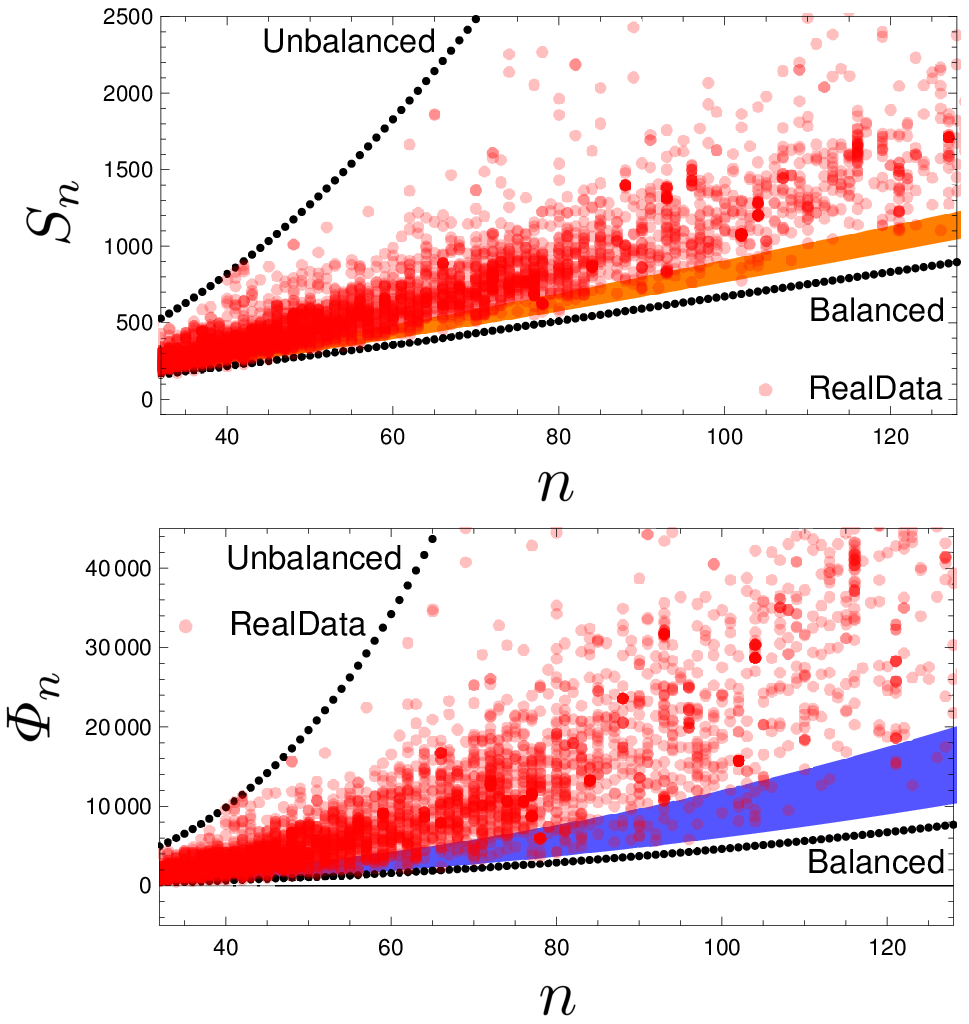} \qquad \includegraphics[width=0.47\textwidth]{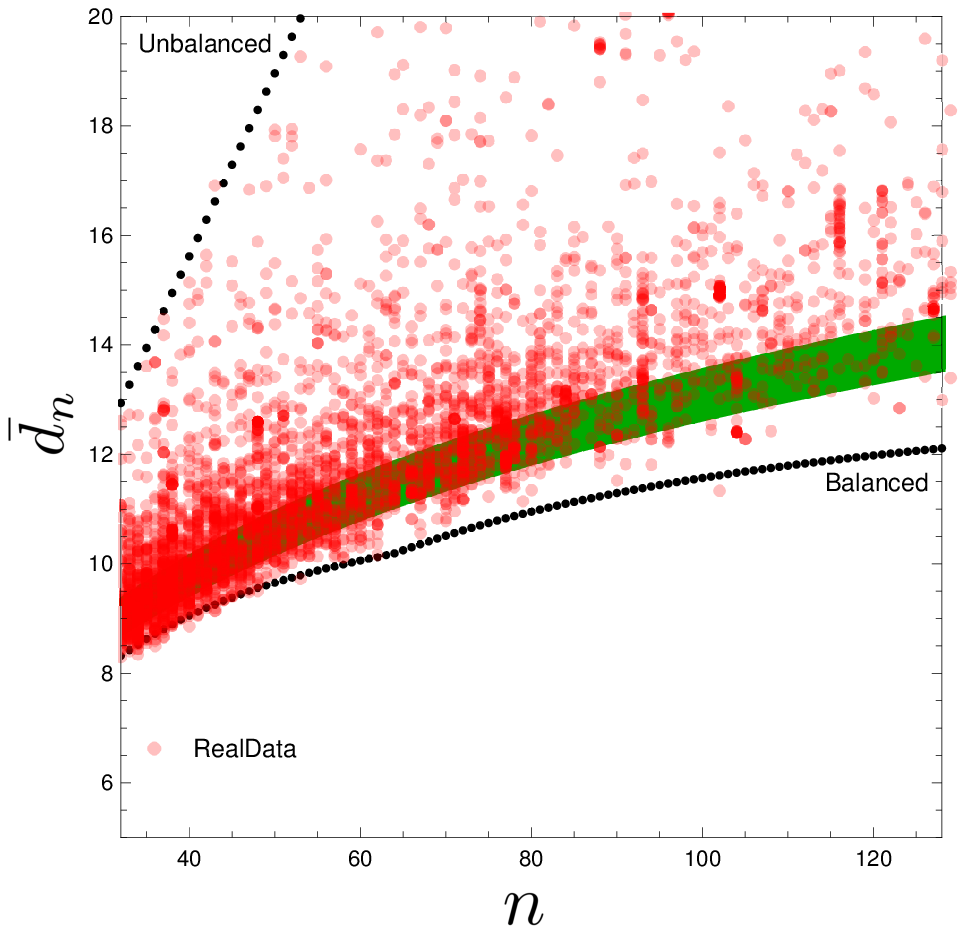}
 \caption{Left Upper Panel: Sackin index for the set of empirical trees. Left Bottom Panel: same for the Total Cophenetic index. Right Panel: same for the APP index.}
 \label{Nnormcasesdata}
\end{figure*}

\begin{figure*}[!htpb]
 \centering
 \includegraphics[width=0.42\textwidth]{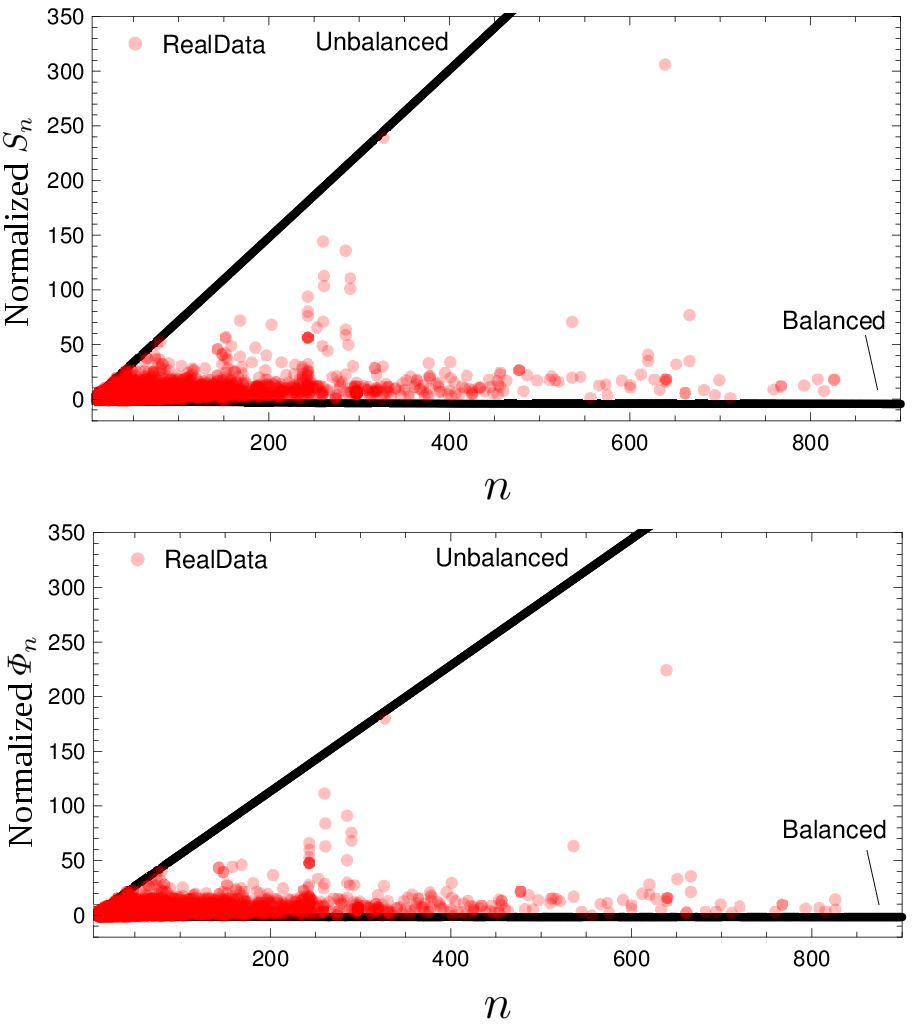} \qquad \includegraphics[width=0.47\textwidth]{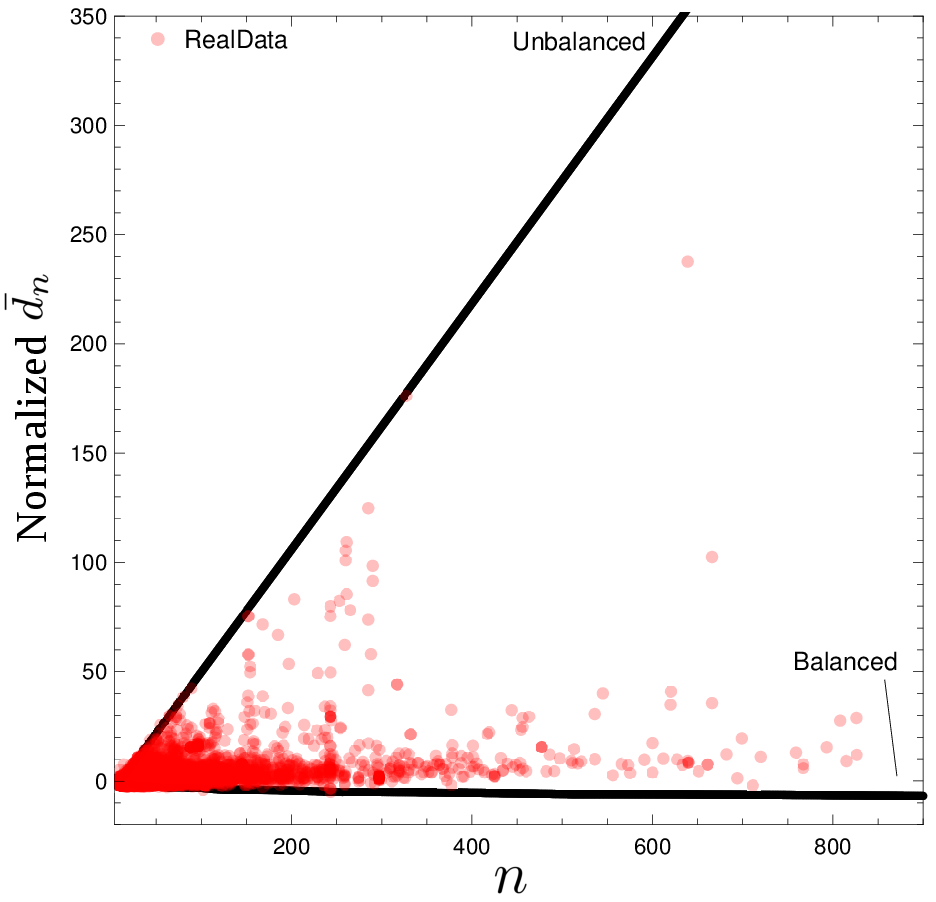}
 \caption{Left Upper Panel: normalized Sackin index for the set of empirical trees. Left Bottom Panel: same for the normalized Total Cophenetic index. Right Panel: same for the normalized APP Index.}
 \label{normcasesdata}
\end{figure*}

We compare the three indices for this set of empirical trees. Figure \ref{Nnormcasesdata} shows a larger concentration of trees near the region delimited by the interval $(\E_Y[\bar{d}_n] - \sigma_Y[\bar{d}_n],\E_Y[\bar{d}_n] + \sigma_Y[\bar{d}_n])$ than for $S_n$ and $\Phi_n$. These intervals are represented by the colored areas. The green area for the APP index captures the distribution of empirical trees better than the Sackin and Total Cophenetic indices. Figure \ref{normcasesdata} shows the same data in terms of the respective normalized indices, where the colored intervals collapse to a narrow window delimited by $(-1,1)$,  since in this case all expected values are null and the variance is unitary. In this case all indices seem to perform equally well.

\section{Conclusions}
\label{conc}

\ \ \ \ The organization of species according to their genetic or phenotypic similarity is usually represented by phylogenetic trees. The shape, branch lengths and distribution of bifurcation points on the tree allows us to reconstruct the evolutionary history of the group.  Because all living species have a single common ancestor, that lived over a billion years ago \citep{ste:2010,the:2010}, the phylogenetic construction can be applied to all species, from small groups that speciated recently to the entire biome. Understanding the structure of these trees is, therefore, key to unravel how life evolved in our planet.\\

In this paper, we have proposed a new index to measure tree balance based on the mean distance between extant species, the Area Per Pair (APP) index. This index has the remarkable property that its variance converges under the Yule Model to a constant value for large phylogenies (\ref{varwass}). Using previously derived expressions for the Sackin and the Total Cophenetic indices, we calculated exact formulas for the APP index for fully unbalanced (\ref{wunb}) and fully balanced (\ref{wbals}) trees.\\

For ensembles of trees generated with the Yule model, both the expected value and variance were calculated exactly  for the APP index, eqs. (\ref{Ew}) and (\ref{varw}). The average of APP index grows slower with tree size when compared to the other two indices. The variance of APP index presents the remarkable property of reaching a constant asymptotic value for large trees -- eq. (\ref{varwass}). For the Sackin and Cophenetic indices, on the other hand, the variance always grows with $n$ -- eqs. (\ref{varSass})  and (\ref{varPhiass}). As the variance of APP index tends to a constant, large trees fall in a narrow range of possible values (Figure \ref{ensemble}). \\

Finally, we calculated the non-normalized and normalized versions of the indices for a large set of empirical phylogenetic trees. We observed a higher concentration of data near the region delimited by the interval $(\E_Y[\bar{d}_n] - \sigma_Y[\bar{d}_n],\E_Y[\bar{d}_n] + \sigma_Y[\bar{d}_n])$ for the APP index (Figures \ref{Nnormcasesdata} and \ref{normcasesdata}). These results suggest that APP is a more suitable index for comparing balance between different (and large) tree sizes.

\bibliographystyle{spbasic}      
\bibliography{(Refs)_2020_JOMB_AraujoLimaMarquittiAguiar_BalanceArea.bib}   

\begin{thebibliography}{23}
\providecommand{\natexlab}[1]{#1}
\providecommand{\url}[1]{{#1}}
\providecommand{\urlprefix}{URL }
\expandafter\ifx\csname urlstyle\endcsname\relax
  \providecommand{\doi}[1]{DOI~\discretionary{}{}{}#1}\else
  \providecommand{\doi}{DOI~\discretionary{}{}{}\begingroup
  \urlstyle{rm}\Url}\fi
\providecommand{\eprint}[2][]{\url{#2}}

\bibitem[{Blum et~al(2006)Blum, Fran{\c{c}}ois, Janson et~al}]{blu:2006}
Blum MG, Fran{\c{c}}ois O, Janson S, et~al (2006) The mean, variance and
  limiting distribution of two statistics sensitive to phylogenetic tree
  balance. The Annals of Applied Probability 16(4):2195--2214

\bibitem[{Cabral et~al(2017)Cabral, Valente, and Hartig}]{cab:2017}
Cabral JS, Valente L, Hartig F (2017) Mechanistic simulation models in
  macroecology and biogeography: state-of-art and prospects. Ecography
  40(2):267 -- 280

\bibitem[{Cardona et~al(2013)Cardona, Mir, and Rossell\'o}]{car:2013}
Cardona G, Mir A, Rossell\'o F (2013) Exact formulas for the variance of
  several balance indices under the yule model. Journal of Mathematical Biology
  67(1):1833 -- 1846

\bibitem[{Colless(1982)}]{col:1982}
Colless DH (1982) Review of phylogenetics: The theory and practice of
  phylogenetic systematics. Systematic Zoology 31(1):100 -- 104

\bibitem[{Coronado et~al(2020)Coronado, Fischer, Herbst, Rosselló, and
  Wicke}]{cor:2020}
Coronado TM, Fischer M, Herbst L, Rosselló F, Wicke K (2020) On the minimum
  value of the colless index and the bifurcating trees that achieve it. arXiv

\bibitem[{Costa et~al(2019)Costa, Lemos-Costa, Marquitti, Fernandes, Ramos,
  Schneider, Martins, and de~Aguiar}]{cos:2019}
Costa CLN, Lemos-Costa P, Marquitti FMD, Fernandes LD, Ramos MF, Schneider DM,
  Martins AB, de~Aguiar MAM (2019) Signatures of microevolutionary processes in
  phylogenetic patterns. Systematic Biology 68(1):131 -- 144

\bibitem[{Felsenstein(2004)}]{fel:2004}
Felsenstein J (2004) {Inferring phylogenies}. Sinauer Associates

\bibitem[{Fischer(2019)}]{fis:2019}
Fischer M (2019) Extremal values of the sackin balance index for rooted binary
  trees. aRxiv

\bibitem[{Harding(1971)}]{har:1971}
Harding EF (1971) The probalities of rooted tree-shapes generated by random
  bifurcation. Advances in Applied Probability 3(1):44 -- 77

\bibitem[{Kirkpatrick and Slatko(1993)}]{kir:1993}
Kirkpatrick M, Slatko M (1993) Search for evolutionary patterns in the shape of
  phylogentic trees. Evolution 47:1171 -- 1181

\bibitem[{Mir et~al(2013)Mir, Rossell\'o, and Rotger}]{mir:2013}
Mir A, Rossell\'o F, Rotger L (2013) A new balance for phylogenetic trees.
  Mathematical Biosciences 241(1):125 -- 136

\bibitem[{Mooers and Heard(1997)}]{moo:1997}
Mooers AO, Heard SB (1997) Inferring evolutionary process from phylogenetic
  tree shape. The quarterly review of Biology 72(1):31--54

\bibitem[{Morlon(2014)}]{mor:2014}
Morlon H (2014) Phylogenetic approaches for studying diversification. Ecology
  Letters 17(4):508 -- 525

\bibitem[{Mulder(2011)}]{mul:2011}
Mulder WH (2011) Probability distributions of ancestries and genealogical
  distances on stochastically generated rooted binary trees. Journal of
  Theoritical Biology 280(1):139 -- 145

\bibitem[{Nee et~al(1992)Nee, Mooers, and Harvey}]{nee:1992}
Nee S, Mooers AO, Harvey PH (1992) Tempo and mode of evolution revealed from
  molecular phylogenies. PNAS 89(17):8322 -- 8326

\bibitem[{Piel et~al(2000)Piel, Donoghue, Sanderson, and
  Netherlands}]{pie:2000}
Piel WH, Donoghue M, Sanderson M, Netherlands L (2000) Treebase: a database of
  phylogenetic information. In: Proceedings of the 2nd International Workshop
  of Species 2000

\bibitem[{{R Core Team}(2017)}]{R:2017}
{R Core Team} (2017) R: A Language and Environment for Statistical Computing. R
  Foundation for Statistical Computing, Vienna, Austria,
  \urlprefix\url{https://www.R-project.org/}

\bibitem[{Sackin(1972)}]{sac:1972}
Sackin M (1972) “good” and “bad” phenograms. Systematic Biology
  21(2):225--226

\bibitem[{Shao and Sokal(1990)}]{sha:1990}
Shao KT, Sokal RR (1990) Tree balance. Systematic Zoology 39(3):266 -- 276

\bibitem[{Steel and McKenzie(2001)}]{ste:2001}
Steel M, McKenzie A (2001) Properties of phylogenetic trees generated by
  yule-type speciation models. Mathematical Biosciences 170(1):91 -- 112

\bibitem[{Steel and Penny(2010)}]{ste:2010}
Steel M, Penny D (2010) Common ancestry put to the test. Nature 465:168 -- 169

\bibitem[{Theobald(2010)}]{the:2010}
Theobald DL (2010) A formal test of the theory of universal common ancestry.
  Nature 465(7295):219 -- 222

\bibitem[{Vos et~al(2012)Vos, Balhoff, Caravas, Holder, Lapp, Maddison,
  Midford, Priyam, Sukumaran, Xia et~al}]{vos:2012}
Vos RA, Balhoff JP, Caravas JA, Holder MT, Lapp H, Maddison WP, Midford PE,
  Priyam A, Sukumaran J, Xia X, et~al (2012) Nexml: rich, extensible, and
  verifiable representation of comparative data and metadata. Systematic
  Biology 61(4):675 -- 689

\end{thebibliography}

\end{document}